\documentclass[]{spie}  
 
\usepackage{amsmath,amsfonts,amssymb}
\usepackage{graphicx}
\usepackage{subcaption}
\usepackage[colorlinks=true, allcolors=blue]{hyperref}
\usepackage{float}
\usepackage{placeins}

\title{Fast, low noise, megapixel detector and readout systems for future X-ray astronomy missions}

\author[a]{Sven Herrmann}
\author[a]{Peter Orel}
\author[a]{Tanmoy Chattopadhyay}
\author[a,b]{Haley R. Stueber}
\author[c]{Jill Juneau}
\author[a,b]{Abigail Y. Pan}
\author[e]{Kevan Donlon}
\author[a]{Declan O'Neill}
\author[c]{Gregory Prigozhin}
\author[c]{Eric D. Miller}
\author[a,d]{R. Glenn Morris}
\author[a]{Tonya L. Peshel}
\author[a]{Artem Poliszczuk}
\author[c]{Beverly LaMarr}
\author[c]{Catherine E. Grant}
\author[e]{Christopher Leitz}
\author[a,b,d]{Steven W. Allen}
\author[f]{Sebastian Albrecht}
\author[c]{Marshall W. Bautz}
\author[e]{Michael Cooper}
\author[f]{Ajay Dakshinamurthy}
\author[c]{Matthew Heine}
\author[f]{Anna Schweingruber}
\author[e]{Keith Warner}
\author[g]{Daniel Wilkins}

\affil[a]{Kavli Institute for Particle Astrophysics and Cosmology, Stanford University, 452 Lomita Mall, Stanford, CA 94305, USA}
\affil[b]{Department of Physics, Stanford University, 382 Via Pueblo Mall, Stanford CA 94305, USA}
\affil[c]{Kavli Institute for Astrophysics and Space Research, Massachusets Institute of Technology, 70 Vassar St, Cambridge, MA 02139, USA}
\affil[d]{SLAC National Accelerator Laboratory, 2575 Sand Hill Road, Menlo Park, CA 94025, USA}
\affil[e]{MIT Lincoln Laboratory, 244 Wood St Building 1324, Lexington, MA 02421, USA}
\affil[f]{Max-Planck Institute for Extraterrestrial Physics, Giessenbachstr. 1, 85748 Garching, Germany}
\affil[g]{The Ohio State University, 4055 McPherson Laboratory, 140 W 18th Ave, Columbus, OH 43210, USA}

\authorinfo{Further author information: Send correspondence to Sven Herrmann \\E-mail: svench@stanford.edu}

\begin{document} 
\maketitle


\begin{abstract}
Next-generation strategic X-ray astronomy missions will require the simultaneous achievement of high angular resolution, large effective collecting area, and wide-field imaging with large-format focal plane detectors. Realizing the associated science objectives—ranging from precision measurements of bright point sources to the detection and characterization of faint diffuse emission—places stringent and, in some cases, competing requirements on detector performance. In particular, high frame rates are necessary to mitigate photon pile-up in observations of bright sources and to reduce contamination from particle-induced background in measurements of low surface brightness structures. At the same time, these instruments must preserve excellent soft X-ray response, which places tight constraints on read noise and on the fidelity of event characterization.

State-of-the-art X-ray charge-coupled devices (CCDs) approach many of the key performance metrics required for these missions, but readout speed remains a primary limitation. Addressing this gap requires readout architectures that scale to high channel count, sustain high pixel throughput, and preserve the low-noise characteristics needed for soft X-ray sensitivity.

\end{abstract}




\section{INTRODUCTION}
\label{sec:intro}  
Next-generation strategic X-ray astronomy missions will require the simultaneous achievement of high angular resolution, large effective collecting area, and wide-field imaging with large-format focal plane detectors. Realizing the associated science objectives—ranging from precision measurements of bright point sources to the detection and characterization of faint diffuse emission—places stringent and, in some cases, competing requirements on detector performance. In particular, high frame rates are necessary to mitigate photon pile-up in observations of bright sources and to reduce contamination from particle-induced background in measurements of low surface brightness structures. At the same time, these instruments must preserve excellent soft X-ray response, which places tight constraints on read noise and on the fidelity of event characterization.
State-of-the-art X-ray charge-coupled devices (CCDs) approach many of the key performance metrics required for these missions, but readout speed remains a primary limitation. Increasing frame rate without compromising noise performance is challenging because higher pixel rates tend to exacerbate noise contributions and power dissipation, and because parasitic capacitances and interconnect complexities become increasingly dominant in discrete, multi-channel readout implementations. Addressing this gap requires readout architectures that scale to high channel count, sustain high pixel throughput, and preserve the low-noise characteristics needed for soft X-ray sensitivity.
To meet these needs, our groups at Stanford, the Massachusetts Institute of Technology (MIT), MIT Lincoln Laboratory (MIT-LL) and the Max Planck Institute for Extraterrestrial Physics (MPE), is pursuing a multi-pronged technology development program.

As one central part we are developing advanced multichannel readout electronics that utilize integrated micro-electronics, waveform sampling, and FPGA based digital signal processing. To support high data throughput, we developed a number of high performance components that can be combined together to support a range of mission profiles. The first of which is a high-speed, low-power, low-noise application-specific integrated circuit (ASIC), denoted as the Multi-Channel Readout Chip (MCRC) that reads out and amplifies the CCD signals. Following amplification, the analog signals are digitized and processed to construct a pixel array, prior to event detection. For this purpose we designed an ADC and FPGA based camera electronics board that enables high-speed, parallelized readout of these CCD channels. The on-board FPGA then performs pre-processing of pixel data into a raw image which then can be further processed for X-ray events extraction and background rejection. 

To scale this architecture toward large-format focal planes, we have expanded our detector system to 16 parallel readout channels and incorporated enhanced diagnostic and debugging capabilities to manage the increased system complexity. Using this upgraded system, we achieved first operation with a large-format 2 Mpixel MIT-LL CCD, providing an initial demonstration of the performance and scalability of the multichannel readout approach.

A second thrust of our program is the continued development of the SiSeRO (Single-electron Sensitive Read Out) technology, a novel readout technology aimed at achieving substantially sub-electron noise at competitive readout speeds. Such performance would directly benefit soft X-ray sensitivity and enable improved event detection and grading, particularly for low-energy photons and low-signal regimes where read noise is a dominant limitation.

Finally, we are developing data-driven methods to further enhance detector performance at the system level. In particular, we are investigating artificial intelligence (AI) techniques for improved particle-background screening and for more accurate event characterization.


\FloatBarrier

\section{Application Specific Integrated Circuits for detector readout}
A central technological challenge for next-generation X-ray observatories is the development of readout electronics suited to the large, high-speed X-ray imagers these missions demand. To reach the necessary frame rates while maintaining or even surpassing the outstanding noise characteristics of cutting-edge X-ray CCDs, the total pixel readout rate (in MPix/s) must be boosted by a factor of 10 to 100. The primary strategy to accomplish this is to increase the number of readout nodes per imager, while simultaneously raising the speed of each individual node. The most effective way to realize this approach is through the design of dedicated integrated electronics, which offer low parasitic capacitance, a compact form factor, and low power consumption—an essential advantage when a large number of channels must be integrated.

\subsection{VERITAS readout ASIC for Athena WFI}
\label{sec:VERITAS}
\begin{figure}[t!]
    \centering
    \includegraphics[width=.5\linewidth]{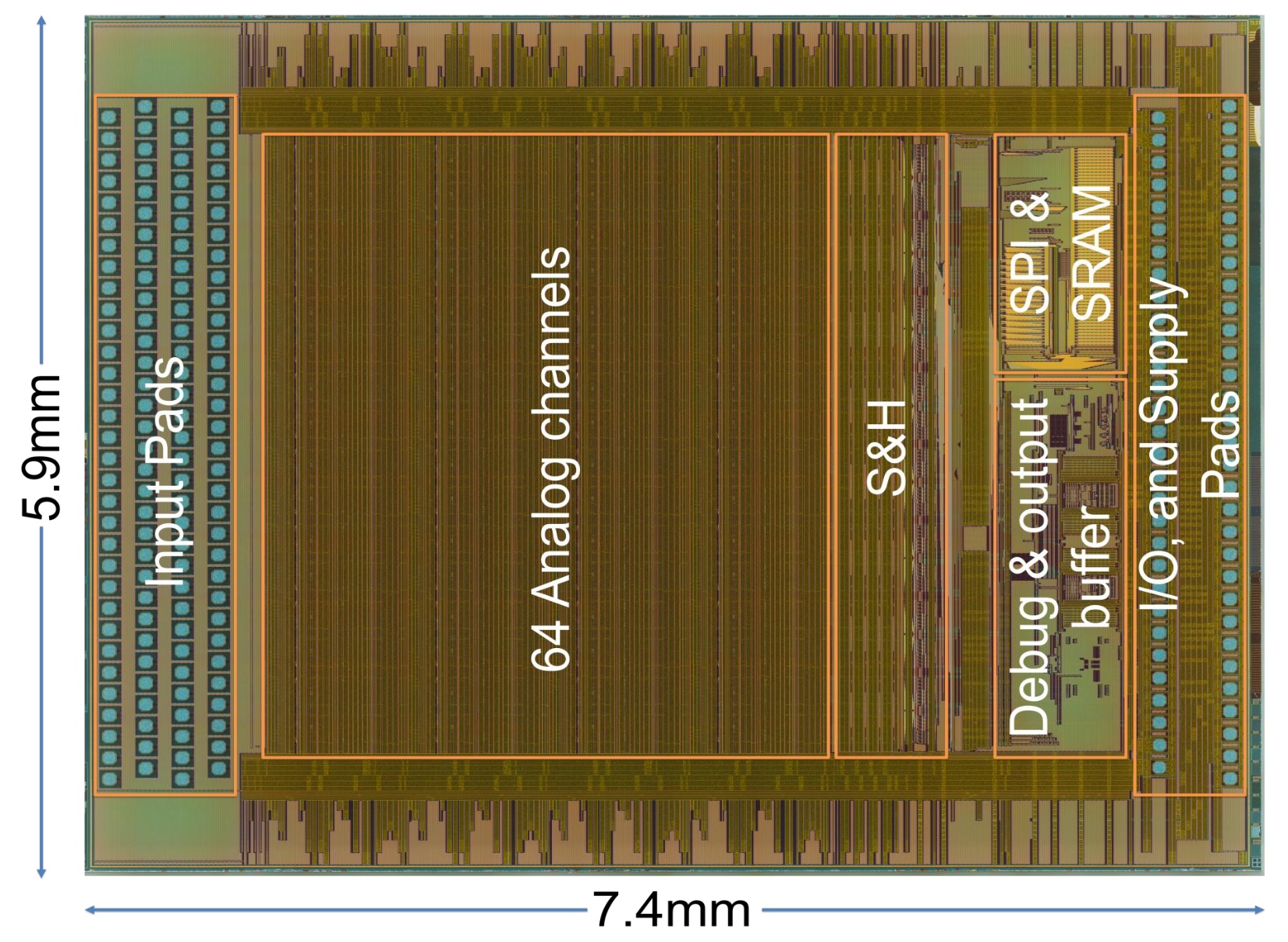}
    \vspace{2mm}
    \caption{Micro-photograph of the fabricated VERITAS 2.3.1 die for the readout of DEPFET detectors. The 64 channel chip provides two sets of input pads (left) for either source follower or drain current readout mode. The center is dominated by the analog amplifier and filtering section, followed by a sample and hold stage with multiplexer. The right edge includes pads for power supply, output buffers and digital I/O.}
    \label{fig:VERITAS231}
\end{figure}

As part of the US contribution to the European Space Agency (ESA) NewAthena mission, our group has pioneered drain current readout architectures (in contrast to conventional source follower voltage readout) for the DEPFET based wide field imager (WFI) and we work collaboratively with the Max Planck Institute for Extraterrestrial Physics on the readout ASIC for the DEPFET detector. Drain readout does not suffer from the settling time limitation as conventional source follower based readouts, which allows for an increase in speed per output but requires highly optimized detector specific input stages on the readout electronics. To address these challenges, the VERITAS\cite{Porro_VERITAS2_2014, Schweingruber_SPIE_2024} (VErsatile Readout based on Integrated Trapezoidal Analog Shapers) readout ASIC, fabricated in a 3.3V 350 nm process technology, has been developed. The chip integrates 64 readout channels operating in parallel for the full column parallel readout of the WFI DEPFET detector, therefore enabling WFI full frame rate of up to 500 frames per second. The VERITAS 2.2 chip is the workhorse for this effort, but due to vendor process technology availability issues the VERITAS architecture needed to be ported to a different process technology vendor (XFAB). The recently manufactured VERITAS 2.3.1 chip, shown in Figure \ref{fig:VERITAS231} is concluding this process technology transition successfully. The performance of the new chip is identical to the previous version, and in addition the new chip includes a number of improvements on power distribution, output buffer and and debugging functionality. Details on VERITAS 2.3.1 can be found in Dakshinamurthy et al. 2026 [\citenum{Ajay2026SPIE}].

\FloatBarrier

\subsection{Multi-channel CCD Readout Chip for mulit-output CCDs }
\label{sec:MCRC}
\begin{figure} [ht!]
   \begin{center}
   \begin{tabular}{c}
   \includegraphics[height=5.2cm]{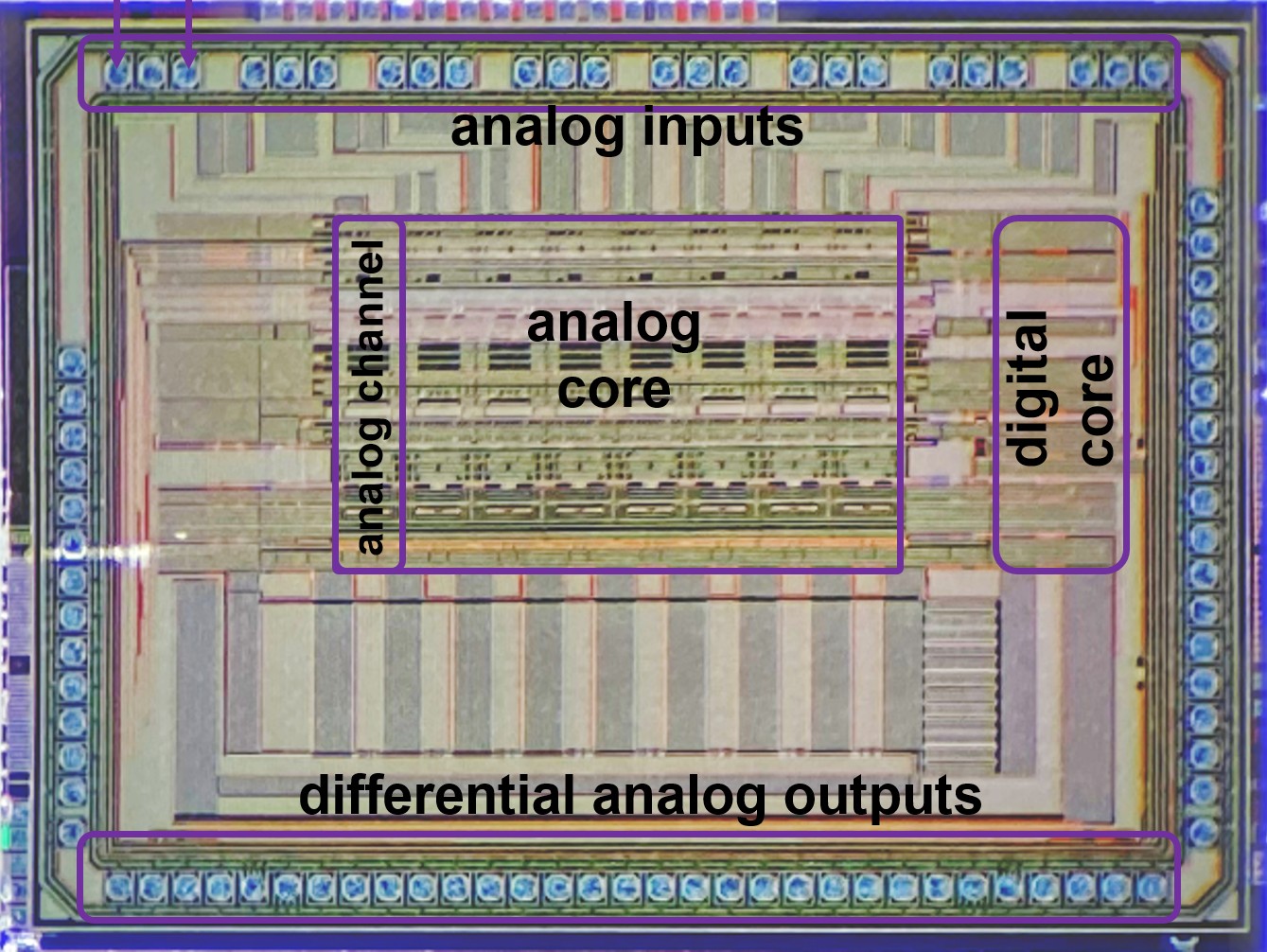}
   \includegraphics[height=5.2cm]{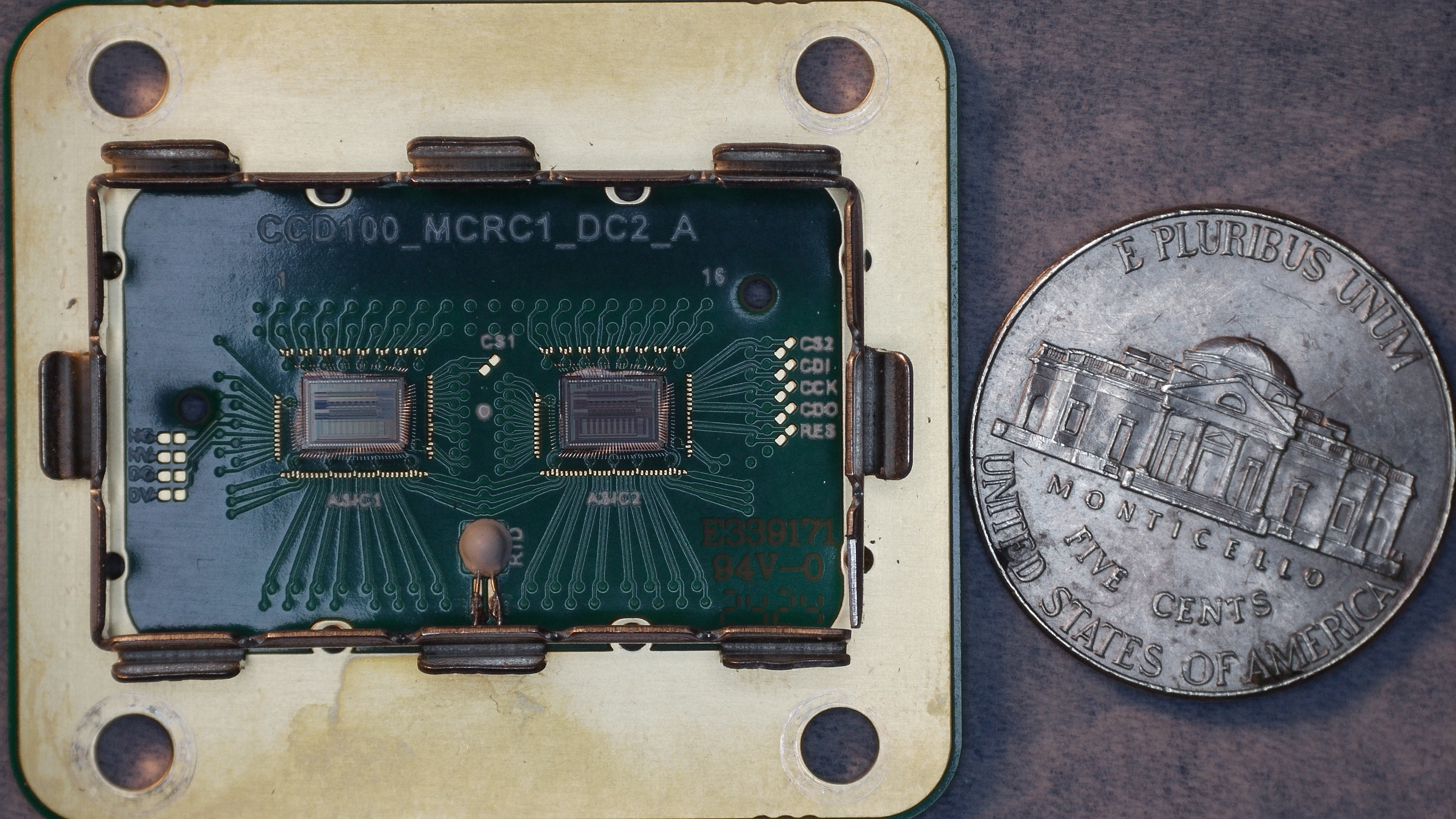}
   \end{tabular}
   \end{center}
   \caption[] 
   { \label{fig:MCRCV1} 
   {\it Left:} Microscope image of an MCRC V1 ASIC readout chip. Dimensions are $4160\,\rm \mu m \times 2900\,\rm \mu m$. The ASIC integrates eight analog readout channels for CCD readout.{\it Right:} 
   Dual MCRC CCID-100 readout board used for 16 channel CCD readout, with a nickel for scale.
}
\end{figure}

The MCRC is an integrated analog readout ASIC, manufactured in a 3.3V 350 nm process technology, designed and optimized for the readout of MIT-LL X-ray CCDs with either JFET or SiSeRO outputs alike. The MCRC-V1 incorporates eight analog readout channels that operate in parallel. Each channel offers two selectable input options that can be connected to the appropriate detector output: a high-impedance source-follower input that includes a programmable current source for biasing the JFET CCD output, or a low-impedance drain readout input that provides a fully integrated solution with a programmable current source, active cascode stage, and current-to-voltage converter to support the SiSeRO operation. Both inputs are followed by a preamplifier that provides selectable gain settings and converts the single-ended input signal into a differential signal, improving immunity to noise and minimizing cross-talk between neighboring channels. The preamplifier output is then fed into a fully differential, unity-gain output buffer capable of directly interfacing with an analog-to-digital converter (ADC) without external buffering, thereby reducing overall system complexity and power consumption. Communication with the ASIC is achieved via a digital serial peripheral interface (SPI) implemented over a low-voltage differential signaling (LVDS) protocol, which is used to program device settings, including internal switch control and digital-to-analog converters (DACs) that generate the bias voltages and currents that set the optimal operating point of the internal analog blocks. The readout chip delivers a large bandwidth, high input and output dynamic ranges, and outperforms the rate and noise capabilities of our best discrete readout solutions for a fraction of the power consumption. The high functional integration of the readout ASIC reduces not only the footprint but also the number of components, significantly simplifying the CCD board design. A microscope image of the MCRC V1 is shown on the left of Figure~\ref{fig:MCRCV1} (dimensions of the chip $\rm 4160\,\mu m \times \rm 2900\,\mu m$) and details of the performance evaluation can be found in Orel et al. 2022 [\citenum{Oreletal2022}] and 2024 [\citenum{porelMCRCspie2024}].

Recently we deployed the MCRC-V1 chip as part of the readout systems at Stanford KIPAC and the MIT Kavli Institute for the characterization of the new 16 channel CCID-100 CCD which requires two of these ASICs operating in parallel. Thus, we built a modular and compact carrier board with two MCRC-V1 chips that can be deployed flexibly. Figure~\ref{fig:MCRCV1} on the right, shows the dual ASIC board that can be plugged into a detector board close to the CCD. We plan to upgrade this board with the next generation of the MCRC (currently in development), that will integrate 16 channels in one chip. Due to the modular nature of the setup, a simple swap of the ASIC board will facilitate the upgrade even in existing CCD setups. 

\section{Multi-output CCDs}
\begin{figure} [ht]
\centering
\begin{subfigure}{0.4\linewidth}
\centering
  \includegraphics[height=6.5cm]{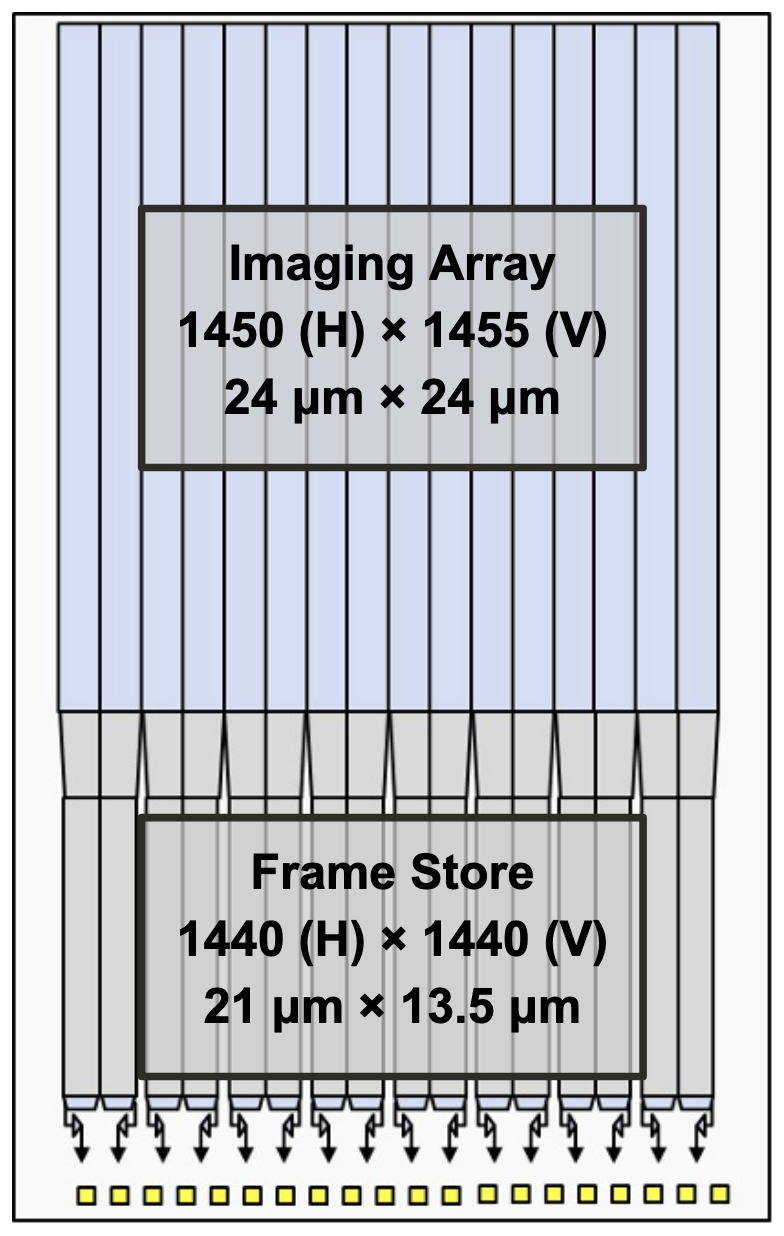}
   \caption[example] 
   { \label{fig:ccid100} CCID-100 architecture.}
   \end{subfigure}
   \begin{subfigure}{0.55\linewidth}
     \centering
    \includegraphics[width=\linewidth]{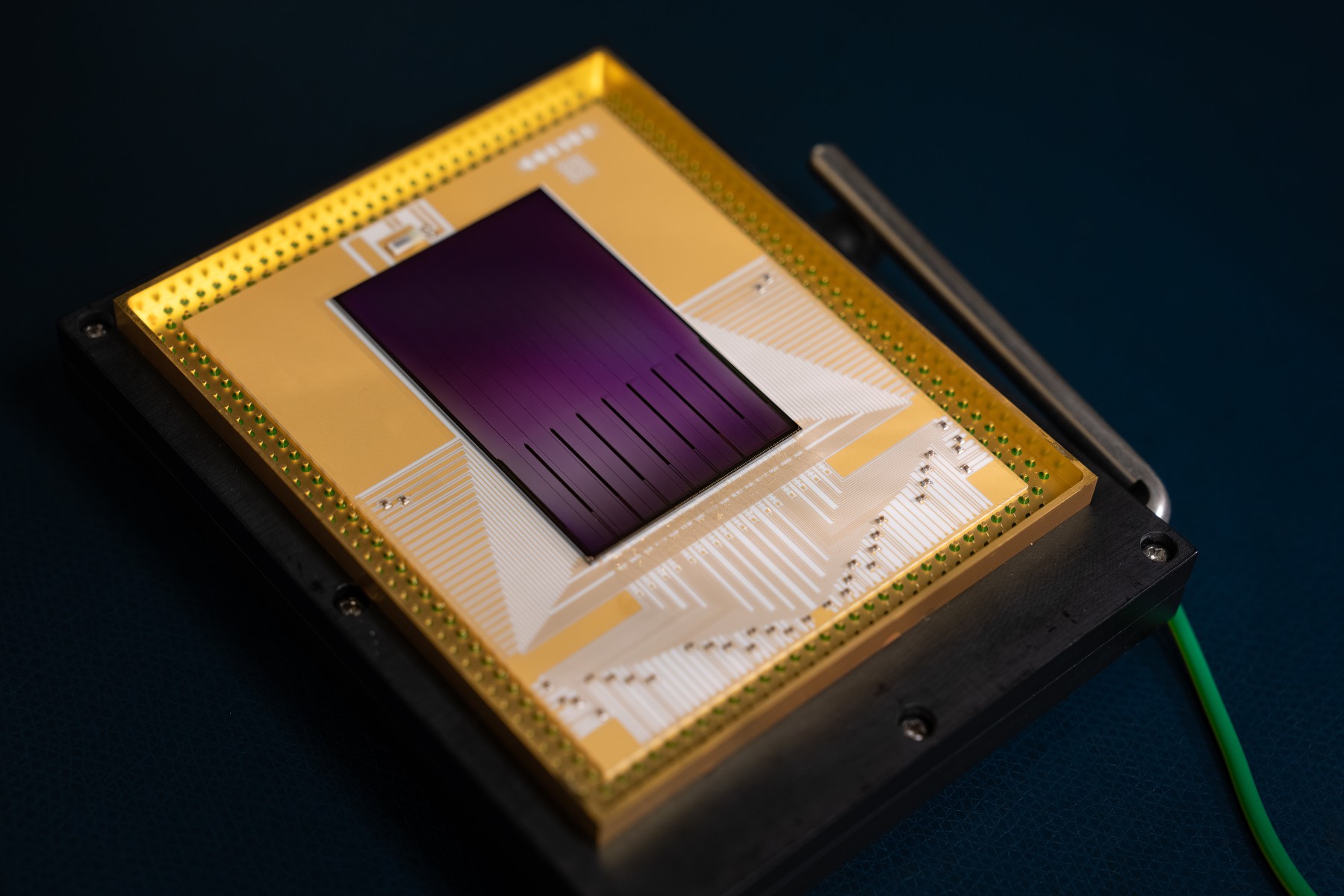}
     \caption{\label{fig:package} CCID-100 mounted in a Kovar package.}
     \end{subfigure}
     \vspace{2mm}
     \caption{The CCID-100 is designed as a large format device with Frame Transfer architecture with an image area of 2.1 Megapixels of 24 $\mu$m size. It features 16 outputs, a considerable step up from legacy devices with 4 outputs, for enhanced readout speed.} 
\end{figure}

\begin{figure} [ht!]
   \vspace{3mm}
   \begin{center}
   \begin{tabular}{c}
   \includegraphics[height=6cm]{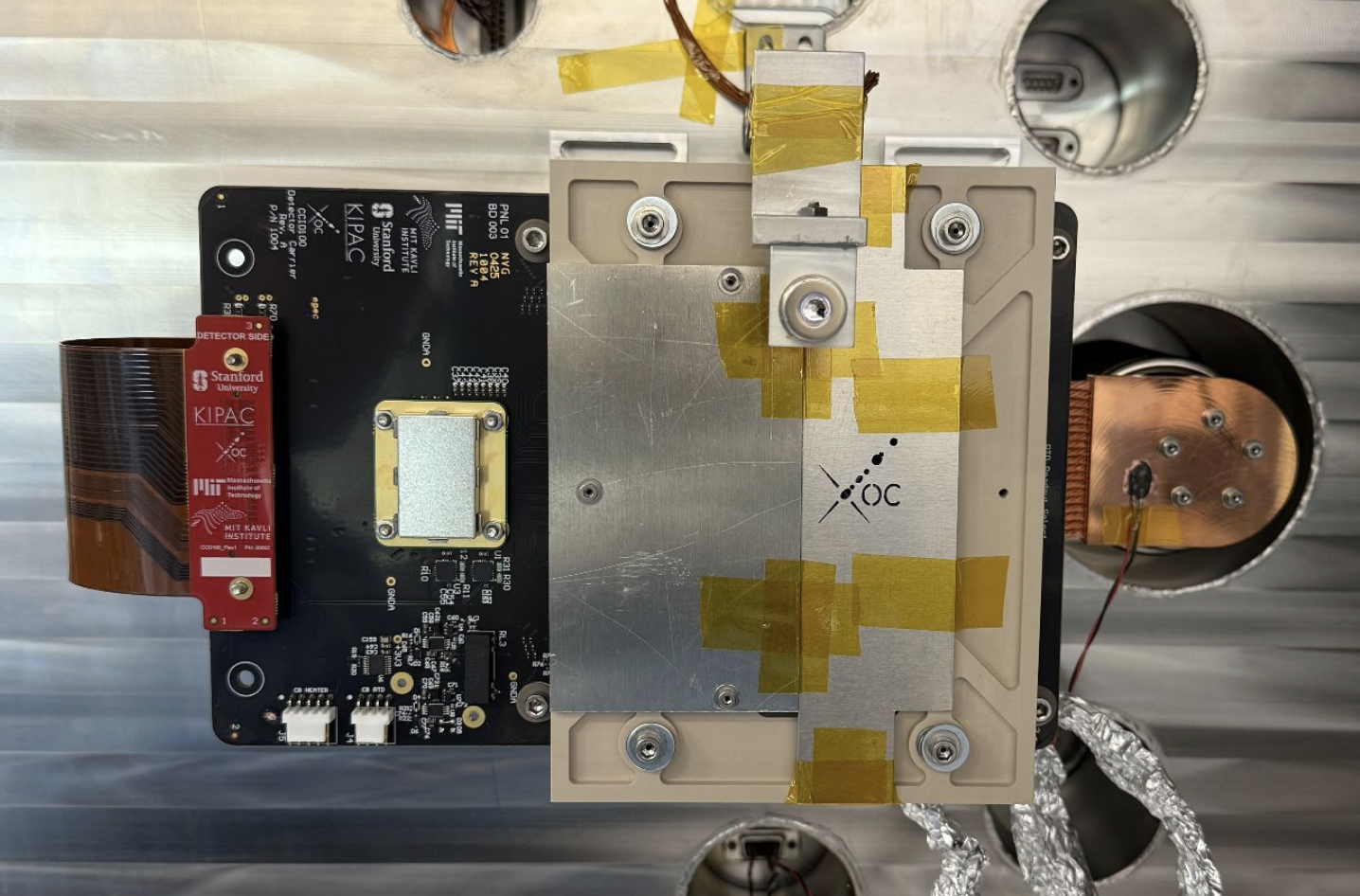}
   \includegraphics[height=6cm]{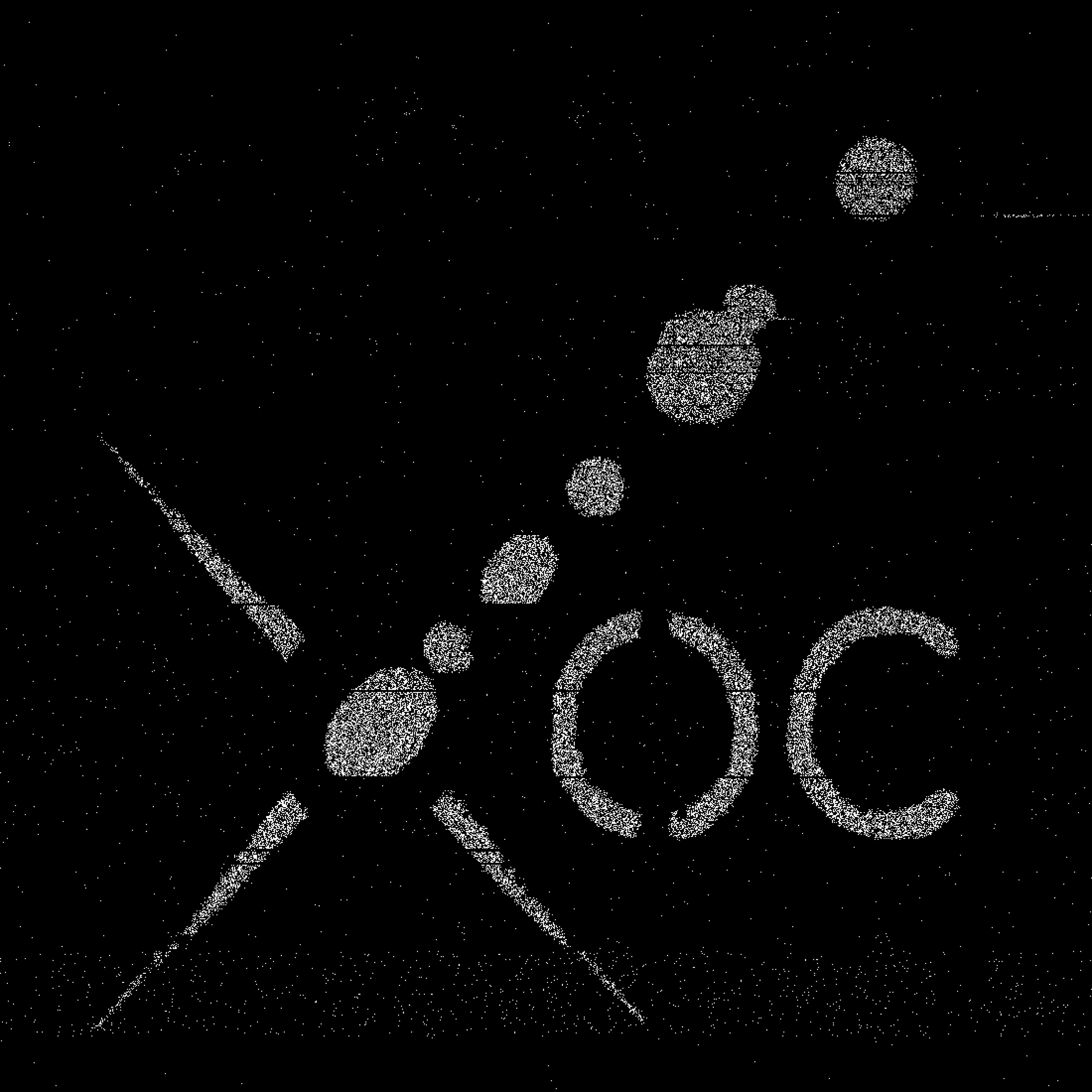}
   \end{tabular}
   \end{center}
   \caption[] 
   { \label{fig:CCID100_board} 
   {\it Left:} Detector mounted in the Gen 1.0 XOC X-ray beamline\cite{stueber2024} with ``XOC'' cutout aluminum cover secured over top of the imaging area.
   {\it Right:} Reconstructed CCID-100 image of single pixel events produced by Titanium (4.5\,keV) fluorescence photons.
}
\end{figure}

MIT Lincoln Laboratory, in partnership with MIT and Stanford University, has been developing a large-format CCD sensor dubbed the CCID-100 for the Advanced X-ray Imaging Satellite (AXIS) mission. While NASA has recently decided not to continue the AXIS probe class mission, the work on the detector technology is continuing as it is expected to be infused into other future X-ray instruments. The CCID-100 is designed as a large format (2.1 Megapixel) device with frame transfer architecture with an image section size of ca. 35 mm $\times$ 35 mm (1450 $\times$ 1455 pixels) and a frame store section (see Fig. \ref{fig:ccid100} ). The CCD includes a total of 16 parallel outputs, resulting in short serial registers of just 90 pixels each. This parallel readout capability results in increased frame rates with a target of up to 20 fps at serial rates of 3.5 MHz. Figure \ref{fig:package} shows a photograph of the CCD device in its Kovar package.
The detector board, which houses the CCD package (by mounting it into an onboard ZIF socket) as well as the dual  ASIC readout module, is shown in Figure \ref{fig:CCID100_board}. It is located inside a vacuum chamber and connects via a potted flex-lead to a commercial STA Archon CCD controller situated outside vacuum. A mask with the groups logo was placed in front of the detector and illuminated with 4.5 keV fluorescence photons. The resulting reconstructed shadow image can be seen in on the right side of \ref{fig:CCID100_board}.
The demonstrated CCD performance of the first four devices, measured at 1 and 2 MHz serial pixel rate is impressive, with excellent device cosmetics, high yield and compatible with the requirements of near future missions: A frame rate of 6.7 frames/sec has been achieved with an average read noise of better than 3 $\mathrm{e}^{-}_{\mathrm{RMS}}$. Further details from the extensive device characterization campaign can be found in Stueber et al. 2026 [\citenum{Stueber2026SPIE}], Prigozhin et al. 2026 [\citenum{Prigozhin2026SPIE}] and LaMarr et al. 2026[\citenum{lamarr26-ccid100}]

\FloatBarrier
\section{Flight Camera Electronics}

\begin{figure}[t!]
    \centering
    \includegraphics[width=.9\linewidth]{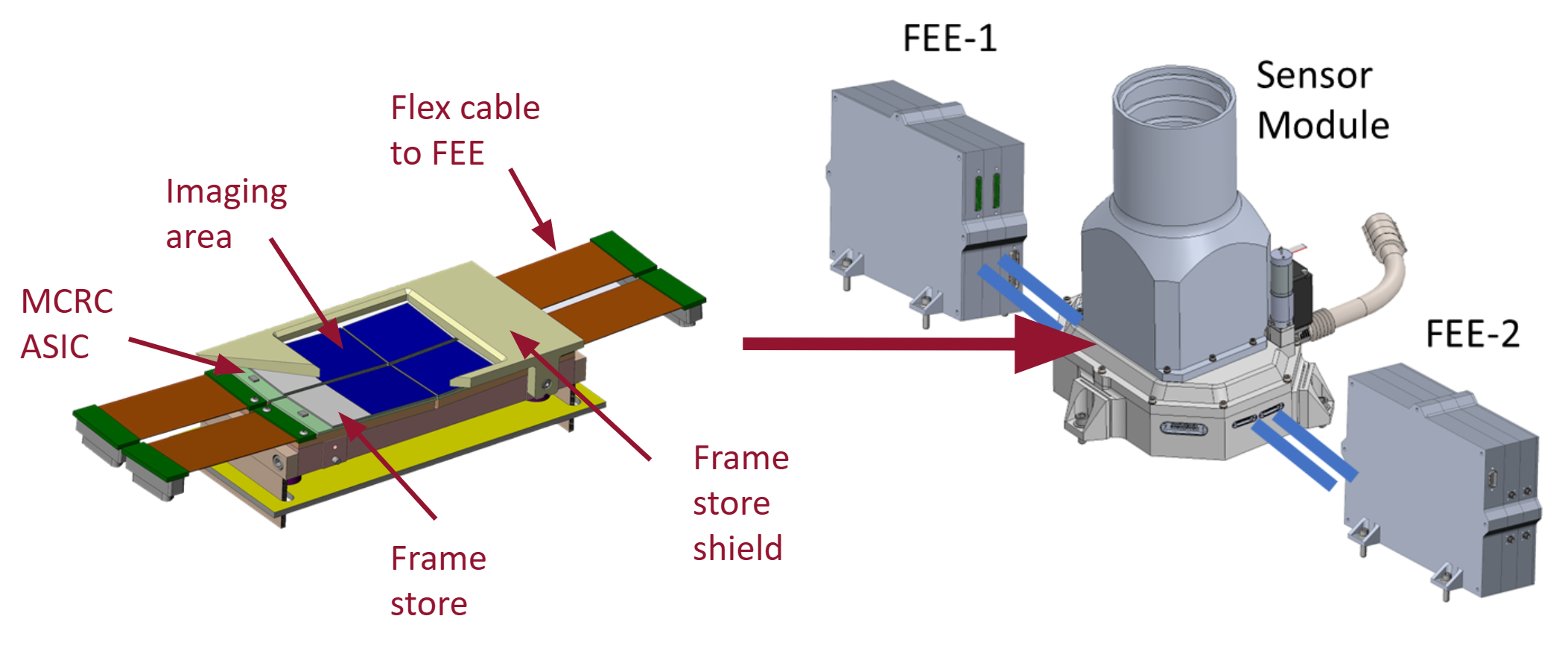}
    \caption{The AXIS focal plane assembly and sensor module is a typical X-ray camera instrument for next generation X-ray missions. The sensor module houses a multi-megapixel focal plane array that is connected via a flex lead to a number of front-end-electronics (FEE) boxes that operate and read out the detectors. The focal plane is cooled to around -100C while the electronics operates around room temperature.}
    \label{fig:axis-camera}
\end{figure}

Another progress based on our TRL advancement for the AXIS X-ray probe class mission\cite{reynolds23} is the development of flight ready camera electronics to support a fully mature X-ray camera instrument\cite{eric25-axiscamera}. The concept of such an instrument is shown in Figure \ref{fig:axis-camera}: a large 8 Megapixel X-ray CCD array is read out with the MCRC ASIC, connected via flex leads to a number of front-end-electronics boxes. This architecture is relevant for a number of possible near future X-ray missions across Explorer, Probe and Flaglet programs. 

In the specific case of AXIS, one FEE box includes 2 CCD controller boards that each capture 16 differential channels from the MCRC, compute pixel values, and transmit those values to the back-end electronics over a 1 Gbit Ethernet connection for event reconstruction. The selected ADCs can sample at 200 MSps shared across up to 8 channels; with 4 ADCs, the FEE can sample each channel at 50 MSps. The FPGA processing architecture provides flexibility in the CCD operating modes: the pixel clock frequency and frame rates are flexible and can be chosen according to the desired tradeoff between noise performance and speed. On the high-speed end, for example, the CCD can operate with a 3.5 MHz serial clock at 20 frames per second. While a low noise mode could operate at 1.0 MHz serial clock rate with 5 frames per second and improved noise performance.

\begin{figure} [ht!]
   \begin{center}
   \begin{tabular}{c}
   \includegraphics[height=4.8cm]{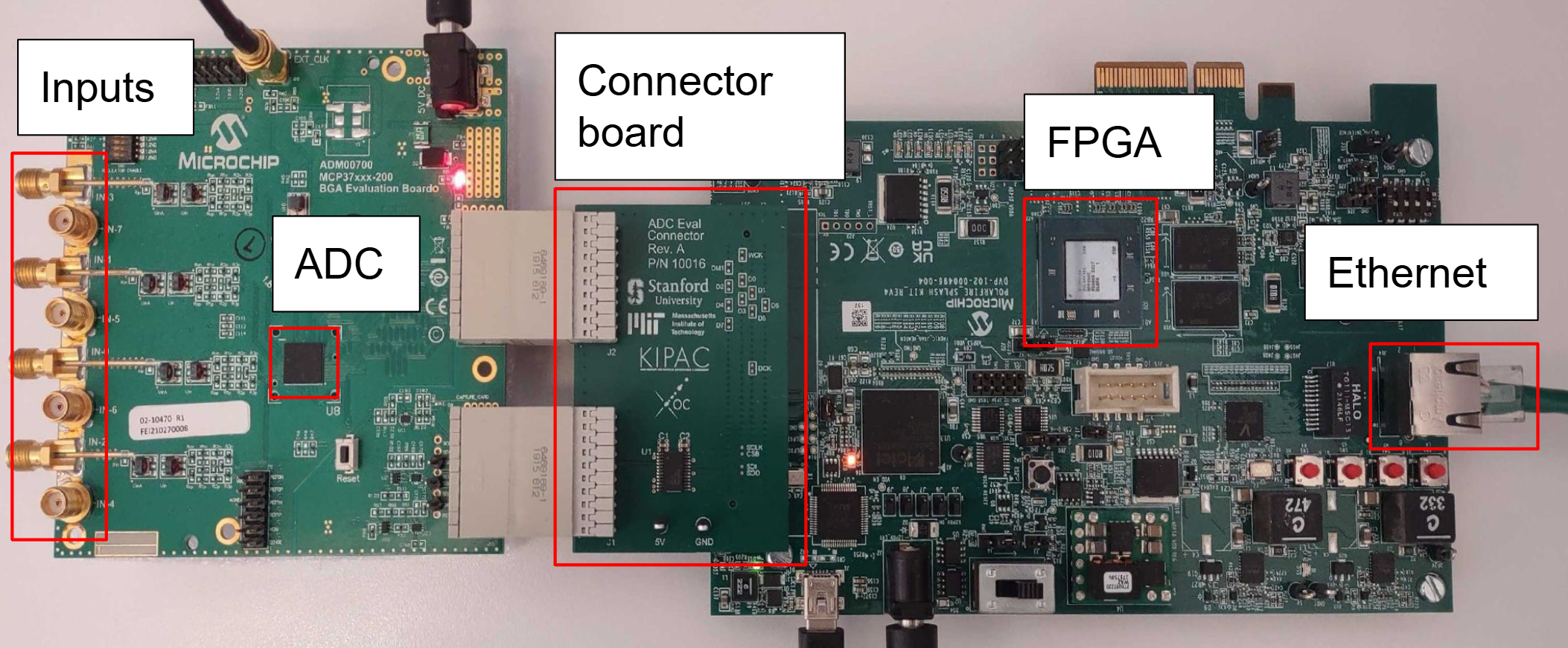}
   \includegraphics[height=5.5cm]{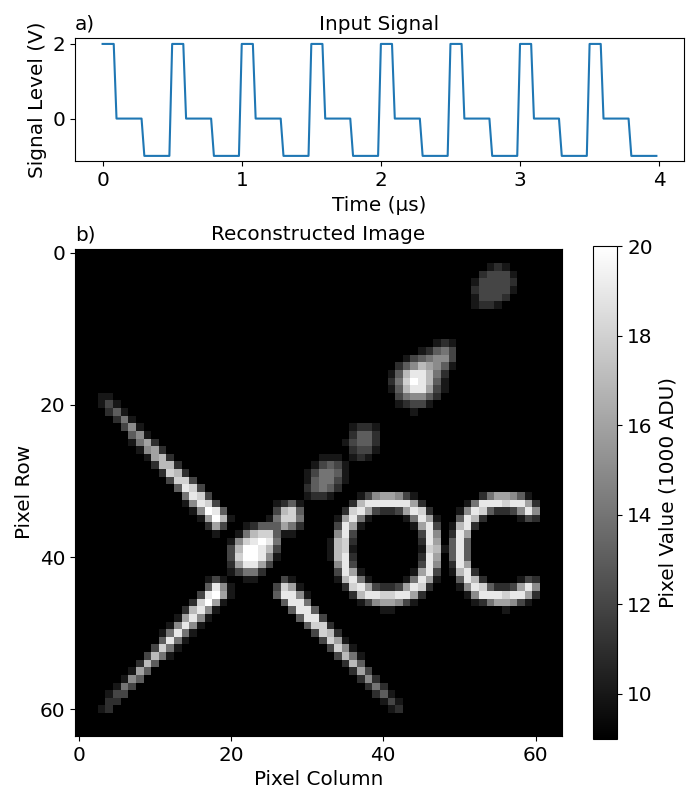}
   \end{tabular}
   \end{center}
   \caption[] 
   { \label{fig:FEE_proto} 
   {\it Left:} FPGA development hardware: on the left is an ADM00700 evaluation board for a MCP37D31-200 ADC and on the right is a PolarFire Splash Kit evaluation board for a MPF300T FPGA. They are linked by a custom connector board that routes the ADC SPI bus and low-voltage differential signaling (LVDS) data and clock signals to the FPGA {\it Right:} Dual MCRC CCID-100 readout board, with a nickel for scale.
}
\end{figure}

To prototype this camera electronics, we built the hardware platform shown in Fig. \ref{fig:FEE_proto}. We identified core components that are available in space qualified versions and used their commercial counterparts for evaluation. The hardware platform is composed of ADC and FPGA evaluation boards bridged by a custom connector board. The developed FPGA firmware captures and processes the ADC datastream of the CCD video waveforms and computes pixel values for a raw image. This raw image can then be processed further in software, running on the FPGA embedded CPU. Data transfer occurs over a 1 Gbit Ethernet connection. Although this prototyping platform includes only a single ADC, the remaining channels are emulated with dummy-data generators so that processing, buffering, and throughput can be tested under representative conditions.
We performed a proof of principle demonstration by driving a CCD-like waveform with known signal levels into one ADC channel and have the electronics perform all the signal processing from waveform to event data and provide a small reconstructed image. The small image on the right side of Figure \ref{fig:FEE_proto} shows the successful result of this demonstration. Detail of this development can be found in O'Neill et al. 2026 [\citenum{ONeil26-axis-fee}]. 

\begin{figure}[t!]
    \centering
    \includegraphics[width=.7\linewidth]{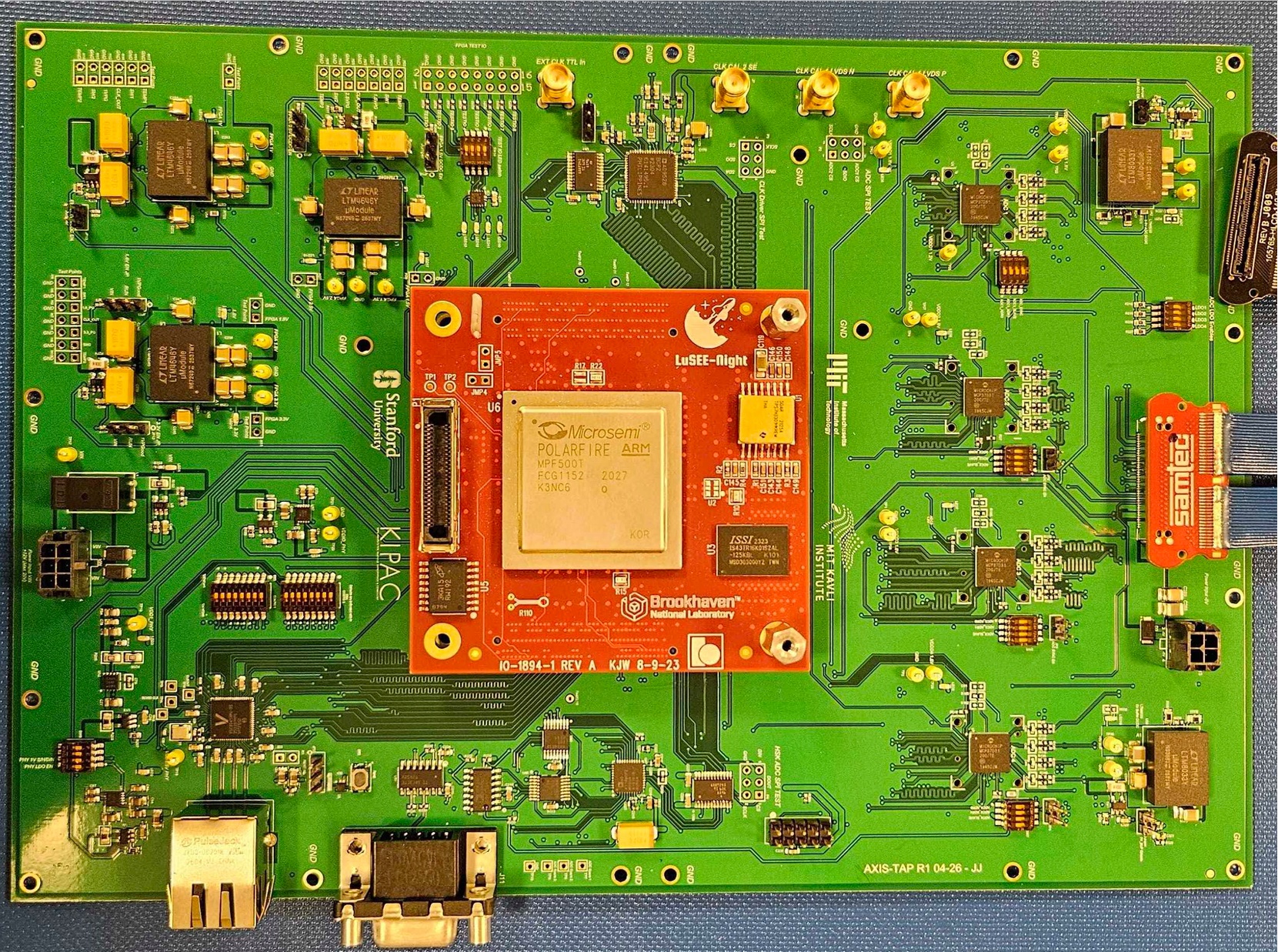}
    \vspace{3mm}
    \caption{AXIS-TAP board that includes a full 16 channel complement of ADCs and a space ready Polarfire 500 FPGA mezzanine board. This board is the next step in TRL advancement for a future X-ray camera; the components have been carefully selected such that direct space qualified alternatives are readily available.}
    \label{fig:FEE_TAP}
\end{figure}

At the same time we also developed together with our colleagues at the Massachusetts Institute of Technology (MIT) a full sized camera electronics board with a full 16 channel complement of ADCs and a space ready Polarfire 500 FPGA mezzanine board to demonstrate scaling of the concept. 
This board was recently fabricated and populated and is now ready to port and scale the prototype firmware to the new FPGA and start testing its functionality. Figure \ref{fig:FEE_TAP} shows a picture of the assembled board; more detail of this development can be found in Juneau et al. 2026 [\citenum{juneau26-axis-tap}].

\FloatBarrier
\section{SiSeRO based detectors}

\begin{figure}[t!]
    \centering
    \includegraphics[width=.5\linewidth]{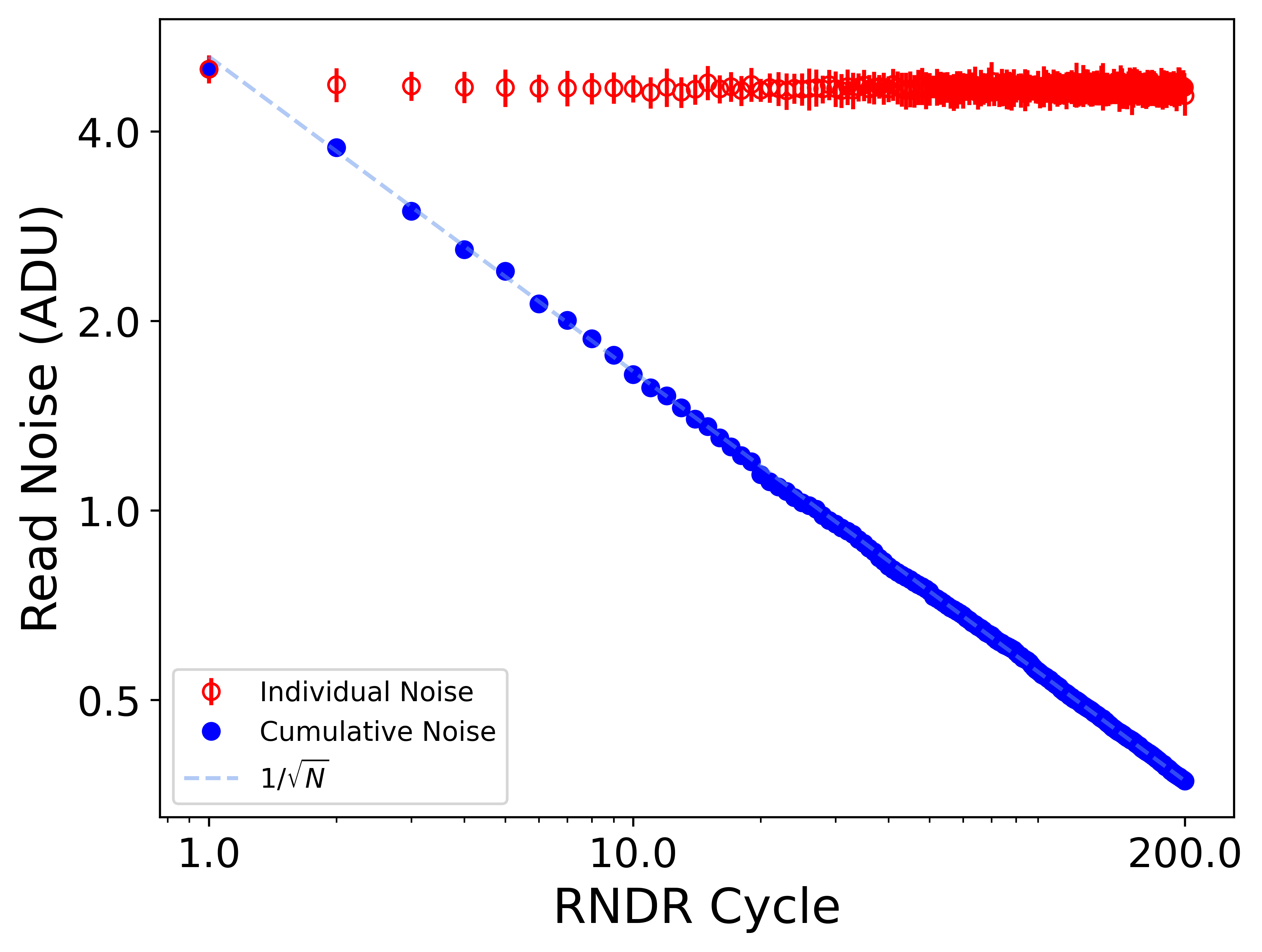}
    \caption{Read noise across 200 RNDR cycles. The red circles show the 200 individual read noise measurements while the blue filled circles are obtained after averaging over the measurements. The cumulative noise measurements follow the expected $1/\mathrm{\sqrt{N}}$ trend, with a final noise of 0.40 $\mathrm{e}^{-}_{\mathrm{RMS}}$ after 200 cycles or about 320 us.}
    \label{fig:RNDR_200}
\end{figure}

A second thrust of our program is continued development of SiSeRO (Single-electron Sensitive Read Out\cite{chattopadhyay22_sisero}), a novel readout technology aimed at achieving substantially sub-electron noise at competitive readout speeds. Such performance would directly benefit soft X-ray sensitivity and enable improved event detection and grading, particularly for low-energy photons and low-signal regimes where read noise is a dominant limitation. The SiSeRO output stage consists of a p-type buried-channel MOSFET fabricated above an internal gate region. Signal electrons stored in the internal gate modify the channel potential and modulate the drain current. One outstanding feature of this device is its capability to support repetitive non-destructive readout (RNDR\cite{chattopadhyay24_rnrdr,chattopadhyay24_rndr_spie}) which allows to measure the exact same charge signal multiple times and reducing the effective read noise by a factor of $\sqrt{\mathrm{N}}$ (where N is the number of RNDR cycles), enabling SiSeROs to achieve sub-electron noise performance as shown in Figure \ref{fig:RNDR_200}.

After the  promising results of first generation SiSeRO devices our team designed and fabricated second generation devices dubbed the CCID-93++ with 3 variants A,B and C. While the A variant uses conventional JFETs as output stage, the B and C variants explore the SiSeRO technology:  

The CCID93++B device has the same 512$\times$512, 8 $\mu$m pixel imaging area plus frame-store as its predecessor the CCID93, but includes 16 SiSeRO outputs in parallel. The 512 serial pixels are divided across 16 SiSeRO outputs, placed at a 256 $\mu$m pitch, which provides a 16$\times$ increase in readout speed and similar noise compared to a single-channel detector. Each detector can also be operating in RNDR mode to reduce the read noise for an increase in readout time. 

The CCID93++C device is similar in pixel format to the CCID93++B device, but here all pixels are read out through a cluster of 16 SiSeRO outputs in series. This is analogous to the multi-amplifier sensor (MAS)\cite{Lapi2024_skipperMAS} CCD.
Each charge packet is passed non-destructively through all 16 outputs during readout and as such every charge packet is measured 16 times by different SiSeRO transistors. Averaging over those measurements results in a $\sqrt{16} = 4\times$ reduction in read noise compared to a single-output device operating at the same frame rate. This device also is capable of additional RNDR modes when needed.  We recently received the first of these new devices from fabrication and further details can be found in Pan et al. 2026 [\citenum{pan26-SiSeRO}]

\begin{figure}[ht!]
   \vspace{2mm}
   \begin{center}
   \begin{tabular}{c}
   \includegraphics[height=5.3cm, keepaspectratio]{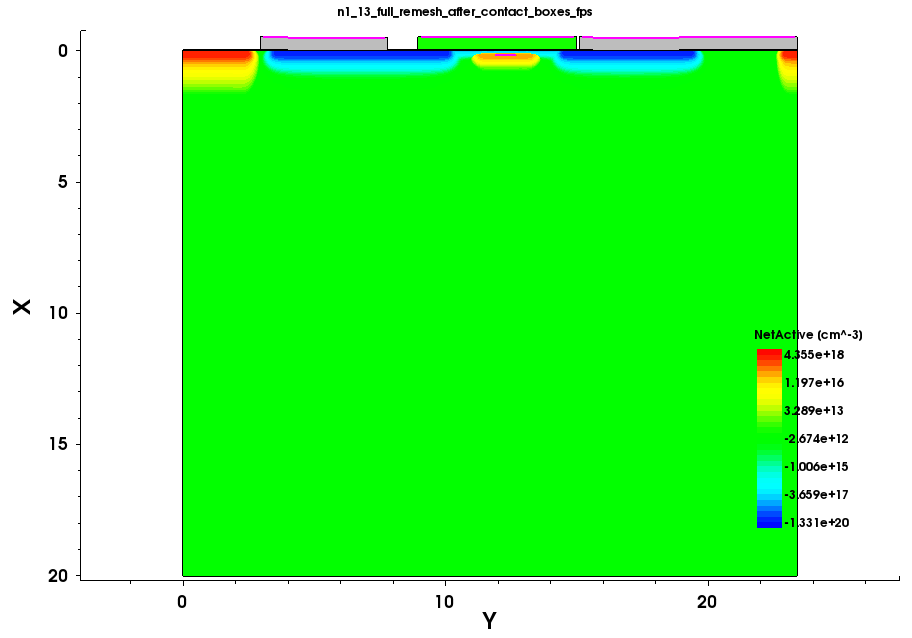}
   \includegraphics[height=5.0cm]{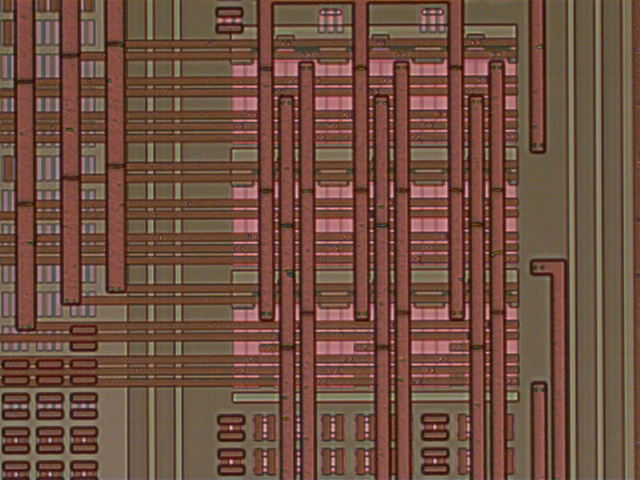}
   \end{tabular}
   \end{center}
   \caption[] 
   { \label{fig:SiSeRO_APS} 
   {\it Left:} {SiSeRO output stage in two-dimensional Sentaurus TCAD model. The model includes p type silicon substrate, a p-MOSFET transistor, source and drain implants, polysilicon gate and an internal gate beneath the p-MOSFET.} {\it Right:} Microphotograph of a recently fabricated 3$\times$3 pixel SiSeRO APS prototype.}
\end{figure} 

Another exciting effort related to SiSeRO devices is our push towards SiSeRO active pixel sensors (APS). SiSeRO devices are manufactured in the same MIT-LL single poly process that is used for the CCDs. Thus we can transition from CCD to APS architectures while maintaining the established advantages of the process technology, such as full depletion and proven entrance windows. Our APS architecture, with a SiSeRO for every pixel, should remedy the two biggest weaknesses of X-ray CCDs, providing the capability to combine full frame, low-noise readout with high speed, region-of-interest readout in the same observation (to avoid pileup and provide precise timing information for bright sources), while minimizing the sensitivity to accumulated radiation induced displacement damage (and subsequent charge transfer inefficiency). Due to the column parallel readout architecture very high frame rates can be achieved (i.e.\ 1000 frames per second for a hypothetical 1 megapixel imager). The same fabrication run that produced the CCID93++ devices also includes a small proof of principle SiSeRO matrix of 3$\times$3 pixels, where every pixel is comprised of two SiSeRO transistors, enabling RNDR readout within every pixel.
In the coming year we will develop a test setup to operate this SiSeRO matrix device and study the basics of its operation. 
In addition, we have also begun to model and simulate the SiSeRO device in Sentaurus TCAD (see Chattopadhyay 2026 et al. [\citenum{chattopadhyay2026_sisero_TCAD}]) and want to reconcile the measurements from the CCID93++ devices and the SiSeRO matrix with the simulations to support the development of a large SiSeRO APS device.

\section{Advanced Algorithms for X-ray event processing}

While technology gaps focus on direct performance parameters for detector instruments, like readout speed and noise, the scientific performance of an instrument is also heavily dependent on the data processing that is applied to the measured values. In the case of X-ray imaging cameras the X-ray event generates a charge cloud in the sensor often across multiple pixels, that needs to be recombined in order to determine the X-ray energy. In addition in a space X-ray telescope the camera will also capture cosmic radiation and secondary X-rays known as background and algorithms try to separate the astrophysical photons from the background. As part of our effort to develop instrumentation for the next generation of space telescopes we are also working on advanced algorithms that leverage machine learning and artificial intelligence approaches - particularly in the two areas of event recombination and background reduction.

\begin{figure}[ht!]
    \begin{center}
    \includegraphics[width=.5\linewidth]{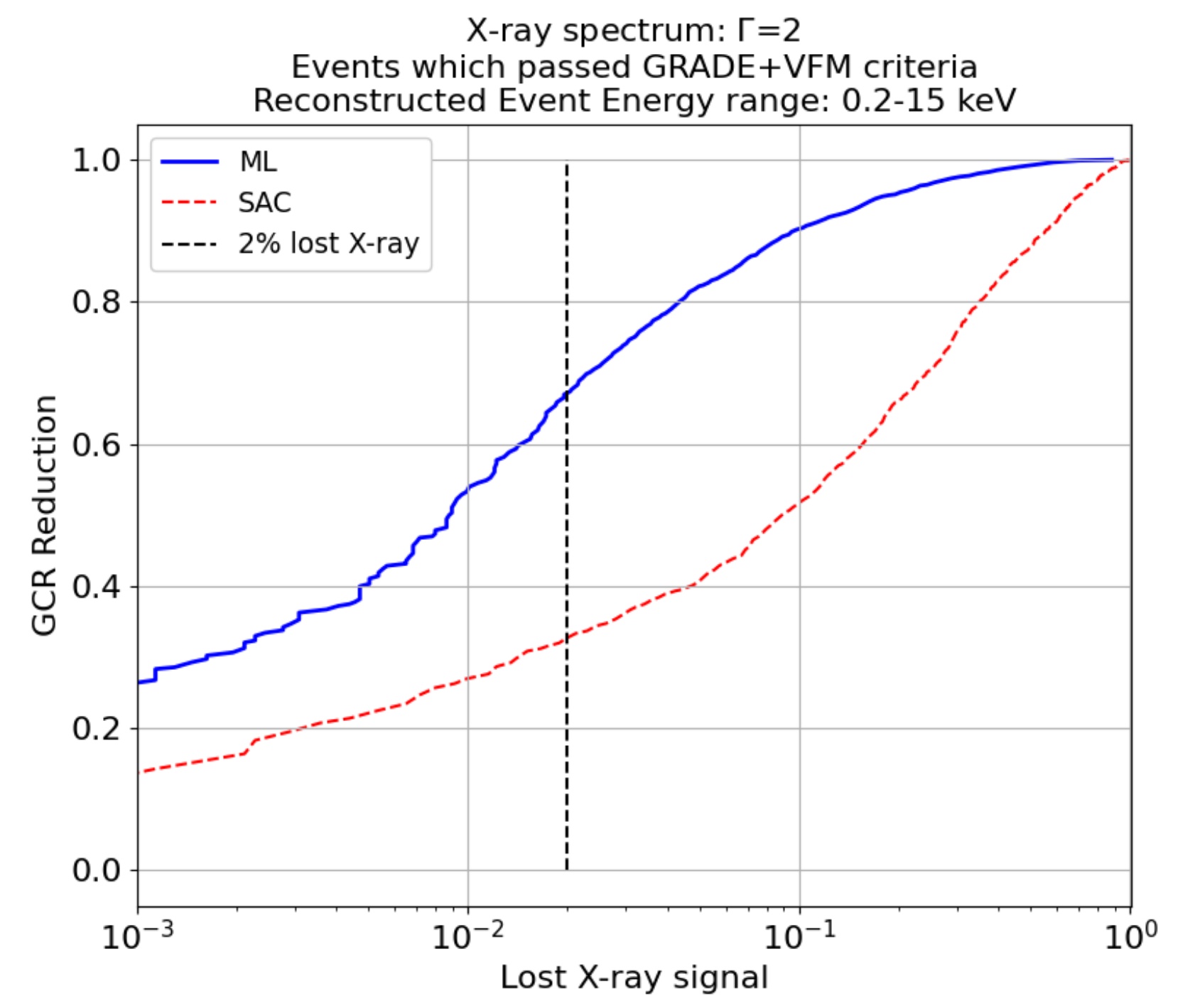}
    \end{center}
    \caption{ROC curve for test data with photons sampled from a different dataset then the training data.  Results are shown for classification of events which pass the Chandra ACIS very faint mode filter as valid events. Blue - our ML model, Red - self-anticoincidence (SAC) model, Black dotted line - acceptable 2\% X-ray loss level.}
    \label{fig:backgroundML}
\end{figure}

Current X-ray space telescopes typically filter cosmic ray background using event reconstruction within 3×3 or 5×5 pixel sliding windows based on designated energy thresholds. While this approach eliminates the majority of cosmic ray signals, the remaining residual background still severely restricts the study of low surface brightness objects and faint, high-redshift X-ray sources. The primary limitation of these traditional filtering algorithms is their inability to account for spatial and energy correlations among various signals within a single frame. In reality, misclassified X-ray events often stem from secondary particles generated when high-energy cosmic rays interact with the detector body. Because signals from both primary and secondary particles frequently co-occur in the same frame, modern machine learning (ML) techniques are uniquely equipped to capture these inherent contextual features.
Our group is developing machine learning (ML)-enhanced methods to optimize cosmic-ray (CR) background reduction \cite{Poliszczuk_SPIE_2024, Wilkins_SPIE_2024} . Tailored specifically for detecting faint, high-redshift sources with the NewATHENA WFI instrument\cite{norbert16_wfi}, our model relies on a two-stage algorithm. It combines weakly supervised event localization via a convolutional neural network with a subsequent random forest classifier, providing a tunable rejection threshold to match specific scientific requirements. By incorporating context awareness absent in traditional algorithms, this method outperforms contemporary alternatives like the self-anticoincidence (SAC) algorithm (SAC)\cite{WFI_bkg_Miller2022JATIS} . Figure \ref{fig:backgroundML} shows the Receiver operating characteristic (ROC) for the two approaches. Our ML framework yields an over 40\% reduction in the CR background—with peak performance ($\geq$ 60\%) achieved for the soft spectra characteristic of high-redshift sources—while restricting X-ray signal loss to just 1–2\%. At the faint limit of the WFI deep survey, this enhanced background mitigation translates directly to over 1.28 improvement in the signal-to-noise ratio (SNR), or over 1.65 reduction in the exposure time required to achieve the target SNR.

The traditional event reconstruction in CCDs (and CMOS detectors) has been based on grading method where the charges spread over an n$\times$n pixel grid are compared with a fixed predefined threshold value (a few times the read noise of the detector). Charges above the threshold are summed over to calculate the total energy of the event. The energy resolution using this method is often poor due to incorrect recognition of the events, particularly at softer X-rays, where the secondary pixel charge can often be less than the threshold value. The incomplete collection of charge results in a low energy tail (therefore poor energy resolution) and reduced gain. While the single pixel event (the charge in the secondary pixels is below the threshold) preserves the best energy resolution, it comes at the expense of reduced number of events (only a few tens of percent of all events depending on the pixel geometry). This effect is worse for small pixel detectors and therefore counterproductive for future high resolution X-ray missions utilizing smaller pixel detectors.

\begin{figure} [ht!]
   \begin{center}
      \includegraphics[height=6cm]{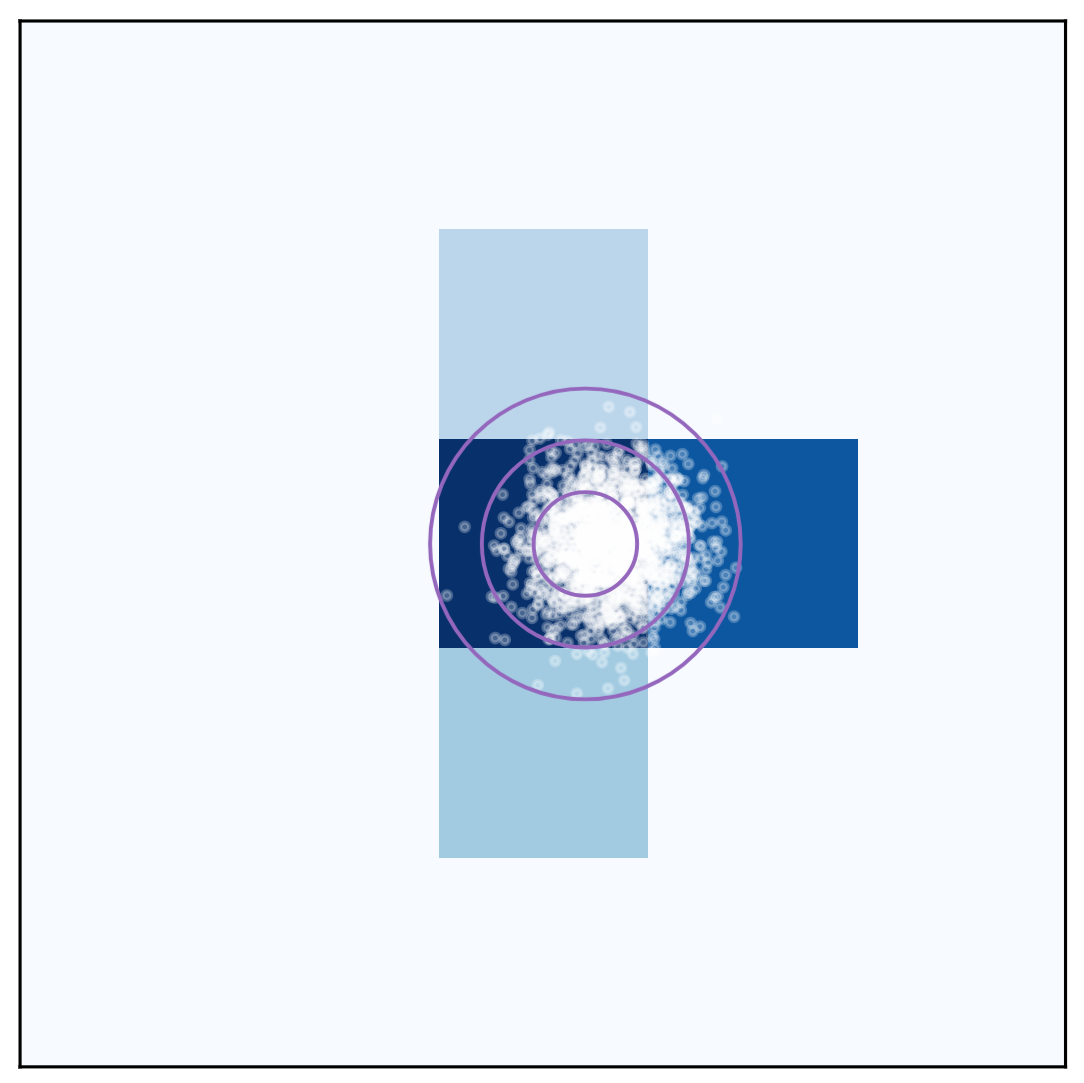}
   \end{center}
   \caption[] 
   { \label{fig:event_recons} 
   X-ray imagers measure the charge packet that is generated by an X-ray photon in the detector. To determine the X-ray energy the charge cloud signal needs to be recombined into the total amount. The white dots indicate the locations of the electrons after diffusion in the pixel grid and the shading color of the pixels represents the value measured in each pixel. The 2D Gaussian fit to the shared pixel charge is shown in purple, the purple lines represent the 1, 2 and $3\sigma$ radii, respectively.  
}
\end{figure}

Our group at Stanford has been working on novel algorithms for event recognition that preserves the number of events without sacrificing the energy resolution \cite{Wilkins_SPIE_2024, BevSPIE2022}. As the charge packet drifts from the interaction point towards the electrodes, it expands thermally due to the thermal random motion of the electrons. The expansion and final charge cloud can be modeled as a Gaussian function. In this method, we model the charge spread across the n$\times$n pixel grid using a two-dimensional (2D) Gaussian function. 
The methodology is shown in Figure \ref{fig:event_recons}. The white dots indicate the locations of the electrons in the pixel grid and the background shading represents the traditional grading method. The 2D Gaussian fit to the shared pixel charge is shown in purple. 

\begin{figure} [ht!]
   \begin{center}
   \begin{tabular}{c}
   \includegraphics[height=6.4cm]{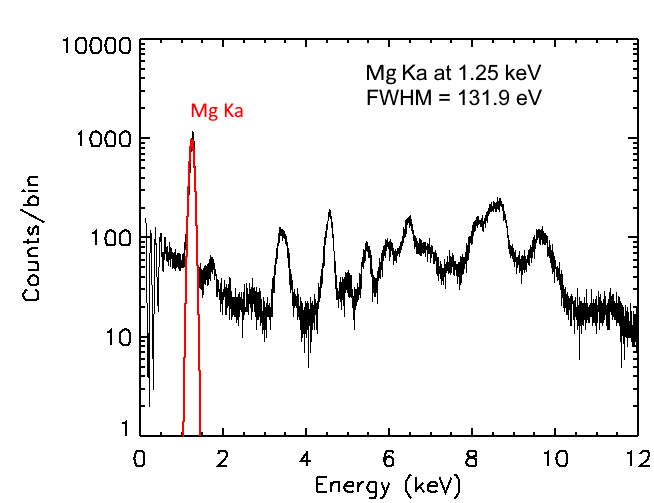}
   \includegraphics[height=5.9cm]{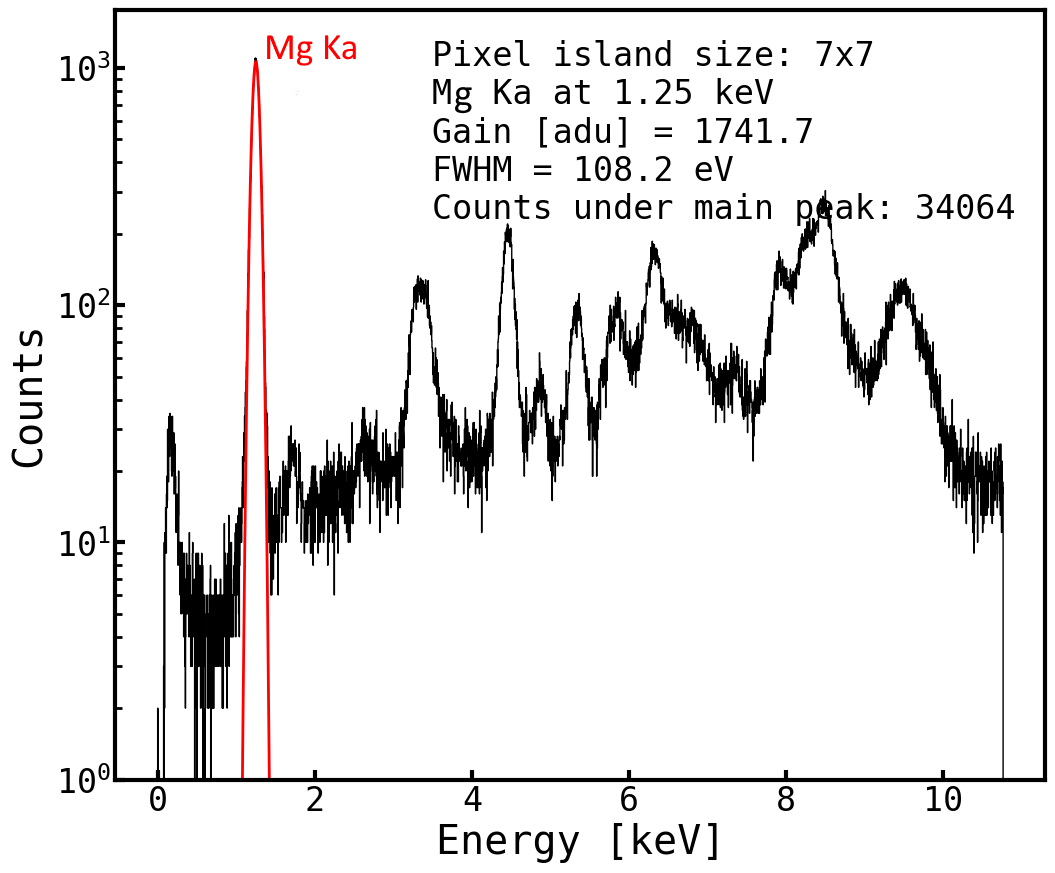}
   \end{tabular}
   \end{center}
   \caption[] 
   { \label{fig:MgK_Events} 
   X-ray spectra of Magnesium K fluorescence for two different processing methods that both show all valid events. {\it Left:} The traditional Chandra ACIS like event processing. {\it Right:} Event processing that utilizes a gaussian fit for the charge cloud. The gaussian fit method improves energy resolution from 131.9 eV to 108.2 eV and improves the low-energy fidelity of the spectrum. 
}
\end{figure}

As a proof of concept of the new method, we used our pJFET based CCID93 detector to collect and reconstruct events for Mg K$\alpha$ photons. Figure \ref{fig:MgK_Events} shows the comparison in the spectra reconstructed using traditional grading method (left) and the new 2D Gaussian method (right).
The 2D Gaussian method results in a significantly better energy resolution without any loss in the total number of valid events.

\FloatBarrier
\section{CONCLUSION}
Recognizing the compelling science drivers, demanding technological requirements and constrained budgets for strategic space missions in the 2030s, our team is working to meet the requirements for cost-effective, high speed, low noise, multi-megapixel NUV-VIS-NIR and X-ray imaging capabilities. Our work leverages the combined benefits of state-of-the-art process technologies for detector fabrication, custom microelectronics for readout, off-the-shelf digital signal processing, and Edge AI algorithms for enhanced data cleaning and event reconstruction within single camera instruments.
Large, 2.1 MPixel X-ray CCDs with excellent device cosmetics and high yield have demonstrated compatibility with the requirements of near future missions, and faster, more capable variants are in evaluation. 
The novel SiSeRO CCDs deliver benchmark read noise performance at a given pixel rate, reaching deep into the sub-electron regime at speeds $>$10kpix/s. Our AI algorithms demonstrate considerable reductions in the particle-induced background, and substantial improvements in all-grade event reconstruction (energy resolution) at the soft X-ray energies of primary interest.
Together, these components and technologies are ready to go for a near-term Probe-class X-ray mission while continuing work towards meeting the (demanding) requirements for future NUV-VIS-NIR and X-ray flagships is underway.

\acknowledgments 

This work has been supported by the NASA APEX grant 80GSFC25CA019, APRA grants 80NSSC22K1921, 80NSSC22K0342, 80NSSC25K7957 and SAT grant 80NSSC23K0211.
The Authors would like to thank for the support from the Kavli Institute for Particle Astrophysics and Cosmology (KIPAC).



\end{document}